# Creators Have Difficulty Abandoning Ideas They Generated

Jin Kim [1,3] and George E. Newman [2]

[1] CUHK Business School, The Chinese University of Hong Kong

[2] Rotman School of Management, University of Toronto

[3] The Institute of Management Research, Seoul National University

## Author Note

Jin Kim https://orcid.org/0000-0002-5013-3958

George E. Newman https://orcid.org/0000-0003-0498-6746

Questionnaires, data, analysis code, and additional materials are openly available at the project's Open Science Framework page, https://osf.io/734em/. We have no conflicts of interest to disclose.

Correspondence concerning this article should be addressed to Jin Kim, CUHK Business School, The Chinese University of Hong Kong, Hong Kong SAR, China. Email: jinkim@cuhk.edu.hk; or George Newman, Rotman School of Management, University of Toronto, Toronto, Canada. Email: george.newman@rotman.utoronto.ca

**Abstract**

Creativity researchers often distinguish between two stages of the creative process: generation versus selection. While much is known about the psychology of idea generation (e.g., the factors that lead to a greater number of novel and useful ideas), less is understood about the nature of selection, or how generation and selection interact. Here we investigate how the act of generating ideas may potentially distort the selection process. Using an incentive-compatible paradigm in which pairs of participants reviewed the same ideas and were rewarded for submitting only high-quality ideas, we find that people submit a greater number of lower-quality ideas when selecting among their own ideas than when selecting among another person's ideas (the *Creative Endowment Effect*). This effect generalizes across three tasks in two domains and is resistant to an informational intervention (i.e., explicitly telling people about the effect). However, having participants revisit their ideas several months later increases their selectivity. The broader implications for individuals and organizations are discussed.

## Research Transparency Statement

### General Disclosures

Informed consent: We obtained informed consent from all participants in each of our studies; no minors were recruited for any of our studies. Conflicts of interest: We have no conflicts of interest to disclose. Funding: The authors received no funding for this work. IRB approval: This research, all experimental protocols, and the bonus payment scheme were approved by and received an exemption determination from the Institutional Review Board at Yale University (IRB#: 0909005737) and conformed to the Declaration of Helsinki. All methods were carried out in accordance with the institution's relevant guidelines and regulations. The payment structure was based on previous related research (Lucas & Nordgren, 2020).

### Studies 1-6

Preregistrations: The hypotheses and methods for Studies 4 and 5 were preregistered prior to data collection at the following addresses, respectively: https://aspredicted.org/W4P_P27; https://aspredicted.org/XLC_WRR. All study materials are available in the Supplementary Information. Data and analysis scripts are publicly available at https://osf.io/734em/

## Statement of Relevance

Using an incentive-compatible paradigm in which participants are incentivized to submit only high-quality ideas, we find that people are less selective when reviewing their own ideas versus another person's ideas. This phenomenon generalizes across tasks and domains and persists even when people are told about the effect. However, having participants revisit their ideas several months later increases their selectivity. These findings make an important and novel theoretical contribution to the creativity literature: While much is known about idea generation, relatively little is known about idea selection and how these two stages may interact. Moreover, our findings suggest fruitful interventions which may improve idea selection in the real world for individuals and organizations.

# The Creative Endowment Effect

Imagine you are working in your office when a moment of creativity strikes. Excitedly, you begin jotting down ideas, one after the other. You rush to your colleague's office, eager to seek their feedback. "Look at this!" you exclaim. "Which of these ideas should I pursue—probably all of them, right?" They carefully review your notes before replying, "I might pursue this one, but I would hold off on the rest." Aghast, you talk through each of the ideas only to realize that they are correct—most of the ideas aren't that great. But then why had those ideas seemed so promising when you first thought of them?

Here we investigate a psychological mechanism that appears to contribute to this phenomenon: People retain more low-quality ideas when reviewing ideas they generated than when reviewing ideas generated by someone else. We call this the *Creative Endowment Effect* (hereafter, *CEE*) and document its existence in multiple domains.

While much is known about how people generate ideas (Björk & Magnusson, 2009; Girotra et al., 2010; Linsey et al., 2005; Shah et al., 2000; Smith, 1998; Toubia, 2006), less is known about the process of idea selection (Faure, 2004)—i.e., how people select among ideas in order to decide which ones to pursue. Individuals and groups who generate more, or more original, ideas do not reliably go on to select the best ones (Faure, 2004; Putman & Paulus, 2009; Rietzschel et al., 2010). Indeed, some research suggests that people prefer to select ideas they personally believe in, or believe should be adopted, rather than the most original or creative ideas generated (Rietzschel et al., 2010). Beyond personal preferences, external factors such as the goals and performance requirements of a problem can also shape the standards people use to evaluate ideas, at times favoring feasibility over long-term originality (Mumford et al., 2002).

Despite this work on idea selection, few studies have directly examined how generation and selection may interact. Accordingly, the present studies investigate an underexamined question regarding how the source of the idea affects its likelihood of being retained during selection. Across six experiments ($N$ = 6,155), we asked participants to engage in a two-stage creativity task. In Stage 1, participants generated ideas. In Stage 2, participants selected among either the ideas that they had generated (Self condition) or ideas that were generated by another participant (Other condition). We utilized a "yoked" design (Nuttin Jr, 1985; Salkind, 2010), such that each set of ideas generated in Stage 1 was reviewed by both the person who generated them and by another participant. All participants were then incentivized to select only the best ideas from the set.

Our paradigm draws inspiration from well-known decision-making effects such as the endowment effect (Kahneman et al., 1990, 1991; Morewedge & Giblin, 2015; Thaler, 1980) and the IKEA effect (Marsh et al., 2018; Norton et al., 2012). In the canonical instantiation of the endowment effect, people value objects they own (e.g., mugs) more than identical objects that they do not own (Kahneman et al., 1990). In the related IKEA effect, participants are asked to bid on either origami cranes that they personally made or origami cranes made by others; participants are willing to pay more for objects that they created (Norton et al., 2012). One can construe the endowment effect and the IKEA effect as instantiations of a broader suite of self-enhancement biases, in which objects and information are assigned higher values by dint of their association with the self (Coulter & Grewal, 2014; Gawronski et al., 2007; cf. Simonsohn, 2011a, 2011b).

With respect to research on creativity per se, a relevant literature examines creators' beliefs about their own creative abilities and performance (Beghetto & Karwowski, 2017). This literature distinguishes between creative self-concept (beliefs that "I am a creative person") and creative self-efficacy (beliefs that "I can perform well, creatively") (Beghetto & Karwowski, 2023). Creative self-efficacy, in particular, can vary with individual and contextual factors that are specific to the task at hand (Bandura, 2012). Related work on the "creative cliff illusion" (Lucas & Nordgren, 2020) demonstrates systematic misjudgments in creators' expectations about how their idea-generation performance will evolve over time as they work: Specifically, creators tend to underestimate how generative and successful they will be in later stages of idea generation. Together, this literature highlights the beliefs people hold about their own creativity and creative performance and shows that such beliefs can be inaccurate. Such beliefs, however, are not the focus of the present research; rather, we examine the process of idea selection—specifically, whether people are more likely to select ideas they generated themselves than ideas generated by others.

Beyond its distinct focus on idea selection, the present research also differs from prior work in the paradigm it employs. In particular, our paradigm differs from those used to study the endowment and IKEA effects in two important ways. First, participants in the IKEA effect studies followed a specific set of instructions. Thus, although the things they made were products of their own labor, they were not particularly creative in any traditional sense. Second, although the IKEA effect and the endowment effect do measure valuation using real-world outcomes, the nature of the valuation is different. In those paradigms, participants indicate how much they value the object. In our paradigm, by contrast, participants are incentivized to select ideas based on their beliefs about how those ideas will be valued by others—a judgment that appears to be particularly difficult and biased when it comes to one's own creations. Indeed, research conducted with circus performers suggests that creators tend to overvalue the success of their own ideas relative to others' (Berg, 2016).

Beyond these distinctions from prior work, the present research examines a common process in organizational settings and everyday life: People brainstorm several ideas (individually or in groups) and then decide which idea(s) to pursue. Thus, the present research informs our understanding of how people engage in this selection process in a way that prior work does not. In the present studies, we go beyond documenting the Creative Endowment Effect by (1) testing whether an informational intervention can attenuate the effect; (2) ruling out alternative explanations, such as miscalibration or idiosyncratic preferences; and (3) demonstrating the moderating role of time. By doing so, we shed light on the psychological mechanisms underlying the effect.

## Results

In Study 1, we utilized a time-lagged yoked design. Participants were asked to brainstorm ideas for how to increase charitable donations to a cancer center. Their task involved two stages: (1) generating ideas for how to increase donations (i.e., the *generation* stage) and (2) reviewing the generated ideas and selecting the best among them (i.e., the *selection* stage). All participants were told that the ideas they selected would be evaluated by an independent panel of judges in terms of their likelihood of increasing donations and that bonus payments would be awarded based on the judges' ratings. Specifically, participants would receive a bonus for every submitted idea that was rated above 3 on a 5-point scale of likelihood of increasing donations, but would lose money for any submitted ideas that were rated below 3; see Supplementary Information, Section 2, for further details regarding the incentive structure.

In the Self condition, participants reviewed and selected the best ideas among those they had generated, whereas in the Other condition, participants (recruited in a second batch) reviewed and selected ideas among those that another participant had generated. Thus, each set of ideas was reviewed by both the person who generated them and by another participant.

As hypothesized, participants submitted more ideas when they selected among their own ideas ($M$ = 4.54, $SD$ = 2.71) than when they selected among others' ideas ($M$ = 3.28, $SD$ = 1.67), Welch's $t(136.5) = 3.62$, $p < .001$, $d = 0.56$. To test whether this expanded selection came at a cost to quality, we recruited 500 participants who reported being affiliated with charity organizations to evaluate the submitted ideas in terms of their likelihood of increasing donations (using the same scale we had provided to participants in the main study). These ratings indicated that the *excess ideas*—those that were retained in the Self condition but not retained in the Other condition—were rated as less likely to increase donations ($M$ = 2.95, $SD$ = 0.46; $n$ = 188) than the remaining submitted ideas ($M$ = 3.09, $SD$ = 0.45; $n$ = 272), Welch's $t(398.1) = -3.34$, $p < .001$, $d = -0.32$.

Study 2 provided a conceptual replication of the effect with the following difference: Participants generated appeals (sentences or slogans) that would encourage people to donate to the same cancer center (see Supplementary Information, Section 3). As in Study 1, participants submitted more ideas when they selected among their own ideas ($M$ = 3.24, $SD$ = 2.21) than when they selected among another participant's ideas ($M$ = 2.39, $SD$ = 1.74), Welch's $t(221.8) = 3.27$, $p = .001$, $d = 0.43$. Moreover, those excess ideas retained in only the Self condition were rated as less likely to encourage donations ($M$ = 2.80, $SD$ = 0.50; $n$ = 214) than the remaining submitted ideas ($M$ = 2.94, $SD$ = 0.46; $n$ = 282), Welch's $t(441.5) = -3.15$, $p = .002$, $d = -0.29$. We also compared the Self and Other conditions to a Control condition in which participants generated and selected ideas in a single stage. This comparison helped determine the direction of the effect: Does a separate selection stage increase selectivity toward others' ideas, or inflate the perceived appeal of one's own? Participants submitted *more* ideas in the Control condition ($M$ = 4.79, $SD$ = 2.84) than in either the Other condition ($M$ = 2.39, $SD$ = 1.74), Welch's $t(182.2) = 7.69$, $p < .001$, $d$ = 1.00, or the Self condition ($M$ = 3.24, $SD$ = 2.21), Welch's $t(209.5) = 4.62$, $p < .001$, $d$ = 0.61. Thus, a separate selection stage increased selectivity in both conditions, but did so more when participants reviewed another participant's ideas than their own—indicating that this difference reflects greater selectivity toward others' ideas rather than an inflated sense of the appeal of one's own ideas.

In Study 3, we tested the CEE with yet another task: generating and selecting humorous captions for a cartoon. Using an established paradigm (Geher et al., 2017; Kim et al., 2013; Lucas & Nordgren, 2020; Sternberg, 2012), all participants were provided with a cartoon and were asked to generate humorous captions. All other aspects of the study were the same as in Study 2 (see Supplementary Information, Section 4).

As in the previous studies, participants submitted more captions when they selected among their own captions ($M$ = 3.79, $SD$ = 1.99) than when they selected among another participant's captions ($M$ = 2.39, $SD$ = 1.36), Welch's $t(196.1) = 6.12$, $p < .001$, $d$ = 0.82. Similarly, ratings from an independent panel of 1,003 participants indicated that the excess captions retained in only the Self condition were rated as less humorous ($M$ = 2.19, $SD$ = 0.46; $n$ = 254) than the remaining submitted captions ($M$ = 2.39, $SD$ = 0.43; $n$ = 268), Welch's $t(514.3) = -5.20$, $p < .001$, $d$ = -0.46. As in Study 2, we also compared the Self and Other conditions to a Control condition in which participants generated and selected captions in a single stage. Participants submitted significantly more captions in the Control condition ($M$ = 4.19, $SD$ = 3.01) than in the Other condition (*M* = 2.39, *SD* = 1.36), Welch's $t(137.7) = 5.53$, $p < .001$, $d = 0.78$, whereas the Control and Self conditions did not significantly differ (*M* = 3.79, *SD* = 1.99), Welch's $t(172.5) = 1.14$, $p = .26$, $d = 0.16$. Thus, the separate selection stage increased selectivity primarily when participants reviewed another participant's captions; unlike in Study 2, the Control and Self conditions did not differ reliably here.

One explanation for the results of Studies 1-3 is that people are miscalibrated about the quality of their own outputs. Participants in the Other condition saw two sets of ideas (the ones they generated, and those generated by another participant), whereas participants in the Self condition only saw one set of ideas. Exposure to more ideas could have led participants in the Other condition to be more selective. In Study 4, we tested this miscalibration account by manipulating whether participants were exposed to examples of top-rated ideas.

Participants were assigned to one of four conditions in a 2 (Idea Source: Self vs. Other) × 2 (Exemplars: Present vs. Absent) between-subjects design. The two "Exemplars Absent" conditions were identical to the Self and Other conditions of Study 3. In the "Exemplars Present" conditions, we provided participants with three captions (randomly selected from a larger pool of 49 captions which qualified for bonus payment) that received the highest average humor ratings in an earlier study (see Supplementary Information, Section 5).

A 2 (Idea Source: Self vs. Other) × 2 (Exemplars: Present vs. Absent) ANOVA with number of captions submitted as the dependent variable revealed only a main effect of Self vs. Other, $F(1, 420) = 35.6$, $p < .001$, $\eta_p^2 = 0.078$. Neither the main effect of Exemplars (Present vs. Absent), $F(1, 420) = 1.2$, $p = .27$, $\eta_p^2 = 0.003$, nor the two-way interaction, $F(1, 420) = 0.016$, $p = .90$, $\eta_p^2 < 0.001$, were significant. In other words, participants submitted more captions in the Self (vs. Other) condition, regardless of whether they saw examples of successful captions prior to their selection.

**Figure 1**

*Quality of Ideas by the Idea Group (Excess Ideas vs. the Remaining Submitted Ideas) and Exemplars Condition*

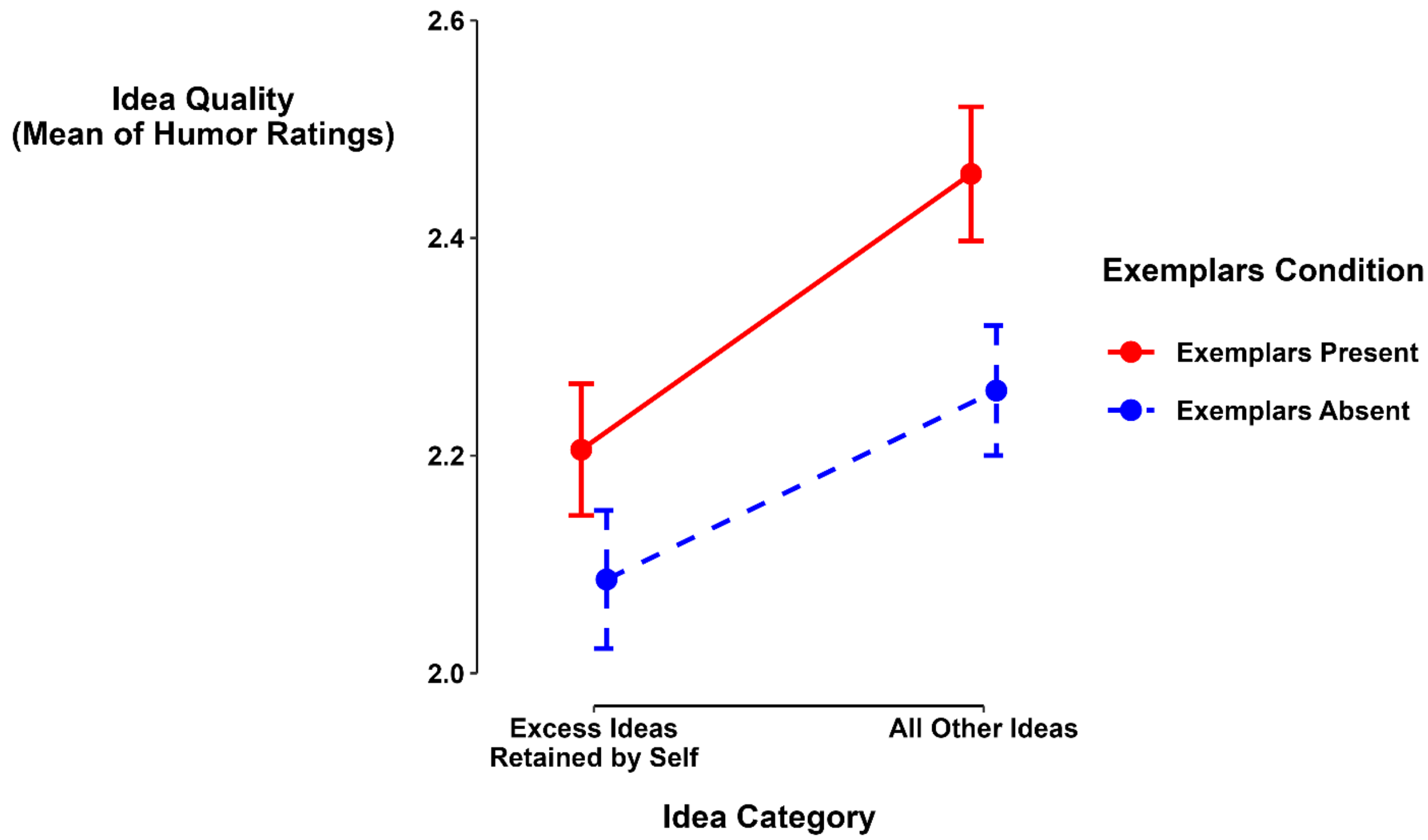

Humor ratings from an independent panel of 865 participants are presented in Fig. 1. Overall, and as intended, the submitted captions were rated as more humorous when participants were given exemplars ($M$ = 2.33, $SD$ = 0.46) than when exemplars were not provided ($M$ = 2.17, $SD$ = 0.48), Welch's $t(856) = 4.84$, $p < .001$, $d = 0.33$. However, the CEE persisted such that the excess captions submitted in only the Self condition were rated as less humorous, both when the exemplars were present ($M_{\text{Excess}} = 2.21$ [$SD_{\text{Excess}} = 0.44$] vs. $M_{\text{Rest}} = 2.46$ [$SD_{\text{Rest}} = 0.44$]; Welch's t[397.3] = −5.79, $p < .001$, $d$ = -0.58) and absent ($M_{\text{Excess}} = 2.09$ [$SD_{\text{Excess}} = 0.49$] vs. $M_{\text{Rest}} = 2.26$ [$SD_{\text{Rest}} = 0.47$]; Welch's $t$[460.5] = –3.93, $p < .001$, d = –0.37). In other words, exposing people to exemplars *did* lead them to select more humorous captions, but that effect was independent of the CEE and did not attenuate the CEE. Thus, the CEE does not appear to arise simply because people are miscalibrated about the quality of their own ideas.

In Study 5, we examined whether the CEE can be attenuated with an informational intervention. The design was similar to Study 4, except that instead of exposing people to exemplars in the two conditions, we told them about the nature of the CEE. Participants were assigned to one of four conditions in a 2 (Idea Source: Self vs. Other) × 2 (Intervention: Present vs. Absent) between-subjects design. The two “Intervention Absent” conditions were again identical to the Self and Other conditions of Study 3. In the two “Intervention Present” conditions, participants read that “research has found that people tend to be overly optimistic when evaluating the quality of their own ideas” and that “people are not critical enough when evaluating their own ideas” (see Supplementary Information, Section 6).

Informing participants about the egocentric bias did not attenuate the CEE. A 2 (Idea Source: Self vs. Other) × 2 (Intervention: Present vs. Absent) ANOVA with number of captions submitted as the dependent variable revealed that only the main effect of Idea Source (Self vs. Other) was significant, $F(1, 362) = 20.0$, $p < .001$, $\eta_p^2 = 0.051$. Neither the main effect of Intervention (Present vs. Absent) nor the two-way interaction was significant, $F$s(1, 362) < 0.56, $p$s > .45, $\eta_p^2$s < 0.002. As in the previous studies, the CEE emerged regardless of whether participants received the informational intervention.

In Study 6, we test another explanation for the CEE: that people retain more of their own ideas because of idiosyncratic preferences. For example, a participant who enjoys wordplay will tend to generate pun-based captions and to find them genuinely funny—so retaining them reflects their idiosyncratic preferences rather than an inflated evaluation of one’s own ideas. To test this account of idiosyncratic preferences, we recruited participants who had generated cartoon captions in the Self conditions (from previous studies) six or more months prior (hereafter, *T1*). At the time of recruitment for Study 6 (hereafter, *T2*), participants reviewed the list of captions they had previously generated (at T1) and were incentivized to submit only the best ones. Thus, we were able to compare participants' selectivity at T1 versus their selectivity at T2 (six or more months later). If the CEE results from idiosyncratic preferences, then participants would retain roughly the same number of captions at T2 as they did at T1. If, however, there is something about in-the-moment generation that drives the CEE, then participants would become more selective after a substantial delay. Indeed, participants retained fewer of their own ideas when they reviewed them at T2 ($M$ = 2.19, $SD$ = 1.57) versus T1 ($M$ = 3.12, $SD$ = 2.13), $t(102) = -4.25$, $p < .001$, $d_{av} = -0.49$. Moreover, the excess ideas retained at T1 were rated as less humorous ($M$ = 2.02, $SD$ = 0.42) than the other selected ideas ($M$ = 2.15, $SD$ = 0.42), Welch's $t(308.0) = 2.63$, $p = .009$, $d = 0.30$.

## Discussion

Across six studies and three idea-generation tasks, we document a Creative Endowment Effect: People retain more of their own ideas than of another person's, and the additional ideas they retain are of lower quality. Unlike many previous studies on creativity, our studies used an incentive-compatible paradigm: Participants were told that they would earn bonuses according to how others evaluated the ideas they selected. This feature helps to distinguish the selection task in our studies from subjective rating tasks in which participants may inflate their self-report ratings of their own ideas.

We examined several potential explanations for the CEE. First, if the effect stems from people being miscalibrated about the quality of their own ideas, showing them exemplars of successful ideas would attenuate it. But we found no such attenuation: Exemplars improved the overall quality of retained ideas but did not attenuate the Self-Other difference in the number of ideas retained (Study 4). Second, if the effect reflects creators' idiosyncratic preferences—people liking and retaining their ideas that match their own taste—the effect should persist over time, since people's tastes would be relatively stable. But it did not: Participants who revisited their own ideas several months later became more selective, attenuating the CEE (Study 6). Finally, if the effect arises because people fail to recognize that they are less critical of their own ideas, informing them of this tendency would attenuate it. But again we found no such attenuation: Telling participants about the CEE did not make them more selective (Study 5).

The CEE may instead arise from the act of idea generation itself: Producing an idea may inflate, at least temporarily, how good that idea seems. Although future work is needed to illuminate this process, one speculative hypothesis may be that it has something to do with what previous research has referred to as a flow state (Csikszentmihalyi, 1990). For example, the positive affect that emerges during idea generation may spill over into evaluations of the ideas produced, as research suggests that affect elicited during one activity can infuse judgments of a separate target (Forgas, 1995).

The present findings likely capture a general evaluation bias, but the practical implications for organizational creativity may be more complex. Research on practiced creators (e.g., Magee, 2022) suggests that initial ideas are typically either abandoned or substantially transformed through drafting and revision. Relatedly, although the tasks used in this research—generating and selecting fundraising ideas, donation appeals, and cartoon captions—are common in experimental research on creativity, they may differ from creative work in organizations, which often involves iteration, collaboration, and structured review. In practice, creativity may also unfold over a longer time horizon as ideas are developed and refined. The CEE may therefore be mitigated during revision, or it may compound, as the initial pool of retained ideas shapes what gets developed further.

Despite these uncertainties, our findings point to two simple interventions that may improve idea selection: (1) asking another person to evaluate one's ideas, and (2) reviewing one's ideas after a substantial delay—both of which would increase selectivity. We hope that our findings will encourage further research on the Creative Endowment Effect and on the interplay between idea generation and selection.

## Statistical Analyses

We analyzed data from all participants who completed the study. Across studies, our primary analyses included independent-samples *t*-tests (and Welch's *t*-tests), factorial ANOVAs, and regression analyses. Comparisons of experimental conditions were performed on the number of ideas that were retained by each participant. Ratings of idea quality were obtained from separate panels of judges, each of whom rated a randomly-selected subset of the submitted ideas. Those ratings were then averaged across raters to provide a mean idea rating which was used in subsequent analyses. We used the following software and R packages for statistical analyses: R (version 4.5.0) (R Core Team, 2025), data.table (version 1.17.4) (Dowle et al., 2025), ggplot2 (version 4.0.1) (Wickham, 2016), and kim (version 0.6.3) (Kim, 2025).

## Data Availability Statement

Questionnaires, pre-registrations, data, analysis code, and additional materials are openly available at the project's Open Science Framework page, https://osf.io/734em/.

## Funding

The authors received no funding for this work.

**Supplementary Information**

**for**

***Creators Have Difficulty Abandoning Ideas They Generated***

## Table of Contents

**Section 1.** Summary of Participants Across Studies

All participants were recruited from Prolific.com ($N = 6{,}155$). Altogether, these participants had a mean age of 30 and consisted of 30% male, 68% female, and 2% other gender. Approximately 70% of them were White, 54% had a bachelor's degree or higher, and their self-reported median household income was between $50,001 and $60,000. Table S1 below shows the demographic information by study and sample type (participants vs. judges).

**Table S1.** Participant Demographics by Study and Sample Type

| Study | Participants or Judges | N | Age | | | Proportion of Genders | | | % White | % Holding Bachelor's Degree or Higher | Median Household Income (Category) |
|---|---|---|---|---|---|---|---|---|---|---|---|
| | | | Mean | SD | Median | Male | Female | Other | | | |
| 1 | Participants | 166 | 24.1 | 6.1 | 22 | 14% | 84% | 1.8% | 77% | 42% | $50,001 - $60,000 |
| 1 | Judges | 500 | 27.7 | 9.1 | 25 | 17% | 81% | 1.2% | 82% | 63% | $60,001 - $70,000 |
| 2 | Participants | 236 | 27.8 | 9.3 | 25 | 31% | 67% | 2.1% | 60% | 56% | $50,001 - $60,000 |
| 2 | Judges | 1029 | 32.9 | 13.0 | 29 | 34% | 65% | 1.0% | 75% | 65% | $60,001 - $70,000 |
| 3 | Participants | 326 | 25.0 | 8.0 | 22 | 16% | 83% | 1.8% | 66% | 44% | $50,001 - $60,000 |
| 3 | Judges | 1003 | 26.0 | 8.4 | 23 | 17% | 81% | 2.0% | 64% | 47% | $50,001 - $60,000 |
| 4 | Participants | 424 | 29.0 | 10.9 | 25 | 34% | 64% | 2.1% | 67% | 46% | $40,001 - $50,000 |
| 4 | Judges | 865 | 29.8 | 11.1 | 27 | 30% | 68% | 1.8% | 72% | 51% | $50,001 - $60,000 |
| 5 | Participants | 366 | 32.7 | 11.6 | 30 | 45% | 52% | 3.6% | 71% | 62% | $60,001 - $70,000 |
| 5 | Judges | 737 | 29.8 | 10.4 | 27 | 38% | 60% | 2.2% | 65% | 53% | $50,001 - $60,000 |
| 6 | Participants | 103 | 30.5 | 11.1 | 26 | 33% | 65% | 1.9% | 72% | 48% | $50,001 - $60,000 |
| 6 | Judges | 400 | 38.0 | 14.4 | 35 | 50% | 48% | 2.3% | 70% | 56% | $60,001 - $70,000 |
| All | Both | 6,155 | 29.7 | 11.3 | 26 | 30% | 68% | 1.9% | 70% | 54% | $50,001 - $60,000 |

**Section 2.** Study 1 Design and Materials

Study Design: Self vs. Other

[Below, the alternative wordings separated by "/" between square brackets show the wordings used for each of the two conditions: (1) Self and (2) Other conditions (respectively).]

[Study Introduction]

Thank you for participating!

In this study, we would like to get your suggestions for how to increase donations for an organization whose goal is to prevent and cure cancer.

The organization is located in a major city. The organization's current goal is to think of ways to increase charitable donations from individuals. Your task is to generate solutions to the problem of how to increase charitable donations to that organization.

[Task Introduction]

First, you will be provided with information about the organization and will be asked to generate as many ideas as you can for how to increase charitable donations.

Then, you will be asked to [select your best ideas / help us select the best ideas]. Specifically, you will be asked to review the ideas that [you have / another participant in this study has] generated and submit them for evaluation by an independent panel. You can submit as many ideas as you would like for the opportunity to earn a bonus payment.

[Bonus Determination Procedure]

Bonus payments will be calculated as follows:

The ideas you submit will be judged by an independent panel of Prolific participants on a 1-to-5 scale (1 = "Definitely would not increase donations" to 5 = "Definitely would increase donations").

For every idea that you submit which receives an average rating of 3 or higher, you will earn a bonus payment of $0.25. However, for every idea that you submit which receives an average rating below 3, you will lose $0.25 in bonus payment. Once all of the ideas you submit have been rated, your total bonus payment will be calculated. Note: Any negative totals will result in a $0 bonus.

Only the ideas that you submit are eligible to contribute to your bonus payment. Accordingly,

whereas you can submit as many of your ideas as you would like, you should only submit the ideas which you think are likely to be rated favorably by the panel.
[The Quiz About the Task and Incentives]

- In this task, is there an opportunity to earn a bonus payment? (Yes / No)
- Whose ideas will you be submitting to receive a bonus payment? (My Ideas / Someone else's Ideas)
- The ideas you submit will be rated by a separate and independent panel of Prolific participants. (True / False)
- Only the ideas you submit are eligible for a bonus payment. (True / False)
- If you submit ideas that are rated less favorably by the independent panel, it will decrease your bonus payment. (True / False)

[Main Task, Stage 1: Idea Generation]

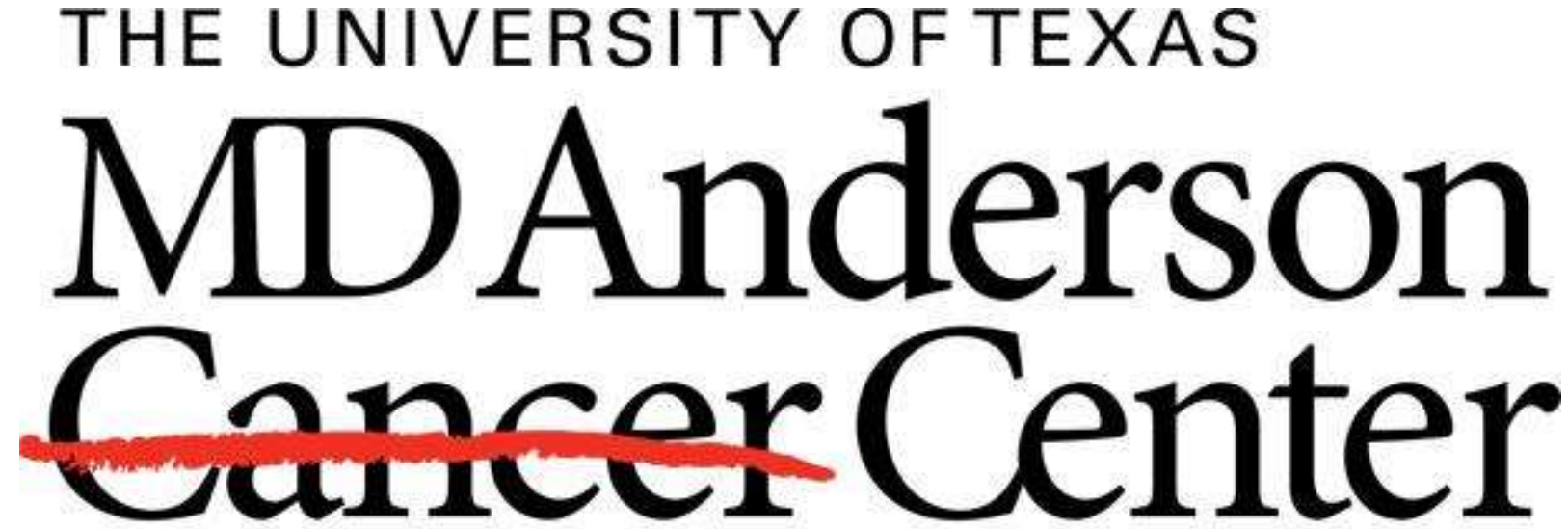

Making Cancer History®

The University of Texas MD Anderson Cancer Center (colloquially MD Anderson Cancer Center) is a comprehensive cancer center in Houston, Texas. It is the largest cancer center in the U.S. and one of the original three comprehensive cancer centers in the country. It is both a degree-granting academic institution and a cancer treatment and research center located at the Texas Medical Center in Houston.

Donations to the MD Anderson Cancer Center make a tremendous difference in the lives of cancer patients by supporting innovative cancer patient care, research, education, and prevention programs. Contributions from individuals dramatically accelerate the pace of converting scientific discoveries into clinical advances to reduce the devastating effects of cancer.

Below, please generate as many ideas as you can for how to increase charitable donations to the MD Anderson Cancer Center. For each idea, try to be as specific as possible:

| | |
|---|---|
| Idea 1: | |
| Idea 2: | |
| Idea 3: | |
| Idea 4: | |
| Idea 5: | |
| Idea 6: | |
| Idea 7: | |
| Idea 8: | |
| Idea 9: | |
| Idea 10: | |

[Main Task, Stage 2: Idea Selection]

Now, we'd like you to review the ideas that [you / another participant] generated.

Please read each of the ideas below and select the idea(s) you wish to submit for evaluation by the independent panel of Prolific participants.

You can also unselect any idea(s) by clicking on them again.

Recall that for every idea that you submit which receives an average rating of 3 or higher, you will earn a bonus payment of $0.25. However, for every idea that you submit which receives an average rating below 3, you will lose $0.25 in bonus payment.

Take as much time as you would like. Once you have selected all of the ideas (and only the ideas) you would like to submit for evaluation, please click the arrow button at the bottom of the page.

[Below, the ideas appeared in buttons that participants could click (highlight) or unclick (unhighlight). All the ideas initially generated (and only these ideas) were shown as ideas that could be selected and submitted. For example, if a participant in the Self condition generated three ideas, all three ideas were presented as three buttons, with Ideas 4-10 not appearing as buttons. An example is shown in Figure S1.]

Idea 1: [the first idea that the participant (Self) or another participant (Other) generated]
Idea 2: [the second idea that the participant (Self) or another participant (Other) generated]
Idea 3: [the third idea that the participant (Self) or another participant (Other) generated]
Idea 4: [the fourth idea that the participant (Self) or another participant (Other) generated]
Idea 5: [the fifth idea that the participant (Self) or another participant (Other) generated]
Idea 6: [the sixth idea that the participant (Self) or another participant (Other) generated]
Idea 7: [the seventh idea that the participant (Self) or another participant (Other) generated]
Idea 8: [the eighth idea that the participant (Self) or another participant (Other) generated]
Idea 9: [the ninth idea that the participant (Self) or another participant (Other) generated]
Idea 10: [the tenth idea that the participant (Self) or another participant (Other) generated]

[Exploratory Items]

You submitted a total of [the number of the given participant's submitted ideas] ideas.

How many of the [the number of the given participant's submitted ideas] ideas do you think will receive an average rating of 3 or higher?

______ idea(s)

You were asked to select among [your / another participant's] ideas. How large a bonus would you have earned if you were instead asked to select among [another participant's / your] ideas?

(Much smaller bonus / Somewhat smaller bonus / Slightly smaller bonus / Same-size bonus / Slightly larger bonus / Somewhat larger bonus / Much larger bonus)

[Demographic Information]

- How old are you? ____
- What is your gender? (Male / Female / Other: ______)
- What is your race or ethnicity? (White / Hispanic, Latino, or Spanish / Black or African American / Asian / American Indian or Alaska Native / Middle Eastern or North African / Native Hawaiian or Other Pacific Islander / Some other race or ethnicity)
- What is the highest level of education you have completed? (Did not complete high school / High school graduate / Some college, no degree / Associate's degree / Bachelor's degree / Mater's, Professional, Doctorate degree)
- Approximately, how much is your annual household income? ($0 - $10,000 / $10,001 - $20,000 / … / $140,001 or Above)

**Figure S1.** Idea Selection Stage (Study 1, Self Condition)

Now, we'd like you to review the ideas that you generated.

Please read each of the ideas below and **select the idea(s) you wish to submit for evaluation** by the independent panel of Prolific participants.

You can also unselect any idea(s) by clicking on them again.

Recall that for every idea that you submit which receives an average rating of 3 or higher, you will **earn a bonus payment of $0.25**. However, for every idea that you submit which receives an average rating below 3, you will **lose $0.25 in bonus payment**.

Take as much time as you would like. Once you have selected all of the ideas (and only the ideas) you would like to submit for evaluation, please click the arrow button at the bottom of the page.

**Idea 1:**
Offer an incentive such as a t-shirt for donations above a certain amount.

**Idea 2:**
Show images of the cancer patients who would benefit from the donations.

**Idea 3:**
Have donors' names listed on a plaque in the cancer center.

**Section 3.** Study 2 Design and Materials

Study Design: Self vs. Other vs. Control

[Below, the alternative wordings separated by "/" between square brackets show the wordings used for each of the three conditions: (1) Self, (2) Other, and (3) Control conditions (respectively).]

[Study Introduction]

Thank you for participating!

In this study, we would like for you to come up with "appeals" for charitable donations (1-2 sentences that will encourage people to donate to an organization).

[Task Introduction]

[First, you will / First, you will / You will] read about an organization and generate as many donation appeals for the organization as you can.

[Then, you will be asked to select your best appeals. Specifically, you will be asked to review the appeals that you have generated and submit them for evaluation by an independent panel. You can submit as many appeals as you would like for an opportunity to earn a bonus payment. /

Then, you will be asked to help us select the best appeals. Specifically, you will be asked to review the appeals that another participant in this study has generated and submit them for evaluation by an independent panel. You can submit as many appeals as you would like for an opportunity to earn a bonus payment. /

You will submit these appeals for evaluation by an independent panel. You can submit as many appeals as you would like for an opportunity to earn a bonus payment.]

[Bonus Determination Procedure]

The appeals you submit will be shown to an independent panel of Prolific participants, who will indicate how likely they would be to donate in response to that appeal on a scale of 1 to 5:

1 = definitely would not donate
2 = probably would not donate
3 = possibly donate
4 = probably would donate
5 = definitely would donate

For every appeal that you submit which receives an average rating of 4 or higher, you will earn a bonus payment of $0.25. However, for every appeal that you submit which receives an average rating below 4, you will lose $0.25 in bonus payment. Once all the appeals you submit have been rated, your total bonus payment will be calculated. Note: Any negative totals will result in a $0 bonus.

Only the appeals that you submit are eligible to contribute to your bonus payment. Accordingly, whereas you can submit as many [of your appeals / of the other participant's appeals / appeals] as you would like, you should only submit the appeals which you think are likely to be rated favorably by the panel.

[The Quiz About the Task and Incentives]

- In this task, is there an opportunity to earn a bonus payment? (Yes / No)
- Whose appeals will you be submitting to receive a bonus payment? (My appeals / Someone else's appeals)
- The appeals you submit will be rated by a separate and independent panel of Prolific participants. (True / False)
- Only the appeals you submit are eligible for a bonus payment. (True / False)
- If you submit appeals that are rated less favorably by the independent panel, your bonus payment will be decreased. (True / False)

[Main Task, Stage 1: Idea Generation]

THE UNIVERSITY OF TEXAS
MD Anderson
Cancer Center

Making Cancer History®

The University of Texas MD Anderson Cancer Center (colloquially MD Anderson Cancer Center) is a comprehensive cancer center in Houston, Texas. It is the largest cancer center in the U.S. and one of the original three comprehensive cancer centers in the country. It is both a degree-granting academic institution and a cancer treatment and research center located at the Texas Medical Center in Houston.

Donations to the MD Anderson Cancer Center make a tremendous difference in the lives of cancer patients by supporting innovative cancer patient care, research, education, and prevention programs. Contributions from individuals dramatically accelerate the pace of converting scientific discoveries into clinical advances to reduce the devastating effects of cancer.

Below, please generate as many appeals as you can for donations to the MD Anderson Cancer Center. "Appeals" are 1-2 sentences that will encourage people to donate to the organization.

[At this stage don't worry about how effective each appeal is—just try to generate as many appeals as possible. /

Recall that for every appeal that you submit which receives an average rating of **4 or higher**, you will **earn a bonus payment of $0.25**. However, for every appeal that you submit which receives an average rating **below 4**, you will **lose $0.25 in bonus payment**.

Take as much time as you would like. Once you have entered all the appeals (and only the appeals) you would like to submit for evaluation, please click the arrow button at the bottom of the page.]

Appeal 1:

Appeal 2:

Appeal 3:

Appeal 4:

Appeal 5:

Appeal 6:

Appeal 7:

Appeal 8:

Appeal 9:

Appeal 10:

[Main Task, Stage 2: Idea Selection, Only in the Self and Other Conditions]

Now, we'd like you to review the appeals that [you / another participant] generated.

Please read each of the appeals below and select the appeal(s) you wish to submit for evaluation by the independent panel of Prolific participants.

You can also unselect any appeal(s) by clicking on them again.

Recall that for every appeal that you submit which receives an average rating of 4 or higher, you will earn a bonus payment of $0.25. However, for every appeal that you submit which receives an average rating below 4, you will lose $0.25 in bonus payment.

Take as much time as you would like. Once you have selected all the appeals (and only the appeals) you would like to submit for evaluation, please click the arrow button at the bottom of the page.

[Below, the appeals appeared in buttons that participants could click (highlight) or unclick (unhighlight). All the appeals initially generated (and only these appeals) were shown as appeals that could be selected and submitted. For example, if a participant in the Self condition generated three appeals, all three appeals were presented as three buttons, with Appeals 4-10 not appearing as buttons. Figure S1. illustrates the analogous page for Study 1.]

Appeal 1: [the first appeal that the participant (Self) or another participant (Other) generated]
Appeal 2: [the second appeal that the participant (Self) or another participant (Other) generated]
Appeal 3: [the third appeal that the participant (Self) or another participant (Other) generated]
Appeal 4: [the fourth appeal that the participant (Self) or another participant (Other) generated]
Appeal 5: [the fifth appeal that the participant (Self) or another participant (Other) generated]
Appeal 6: [the sixth appeal that the participant (Self) or another participant (Other) generated]
Appeal 7: [the seventh appeal that the participant (Self) or another participant (Other) generated]
Appeal 8: [the eighth appeal that the participant (Self) or another participant (Other) generated]
Appeal 9: [the ninth appeal that the participant (Self) or another participant (Other) generated]
Appeal 10: [the tenth appeal that the participant (Self) or another participant (Other) generated]

[Exploratory Item 1]

You submitted a total of [the number of the given participant's submitted appeals] appeals.

How many of the [the number of the given participant's submitted appeals] appeals do you think will receive an average rating of 4 or higher?

______ appeal(s)

[Exploratory Item 2, Only in the Self and Other Conditions]

You were asked to select among [your / another participant's] appeals. How large a bonus would you have earned if you were instead asked to select among [another participant's / your] appeals?

(Much smaller bonus / Somewhat smaller bonus / Slightly smaller bonus / Same-size bonus / Slightly larger bonus / Somewhat larger bonus / Much larger bonus)

[Demographic Information]

- How old are you? ____
- What is your gender? (Male / Female / Other: ______)
- What is your race or ethnicity? (White / Hispanic, Latino, or Spanish / Black or African American / Asian / American Indian or Alaska Native / Middle Eastern or North African / Native Hawaiian or Other Pacific Islander / Some other race or ethnicity)
- What is the highest level of education you have completed? (Did not complete high school / High school graduate / Some college, no degree / Associate's degree / Bachelor's degree / Mater's, Professional, Doctorate degree)
- Approximately, how much is your annual household income? ($0 - $10,000 / $10,001 - $20,000 / … / $140,001 or Above)

**Section 4.** Study 3 Design and Materials

Study Design: Self vs. Other vs. Control

[Below, the alternative wordings separated by "/" between square brackets show the wordings used for each of the three conditions: (1) Self, (2) Other, (3) Control conditions (respectively).]

[Study Introduction]

Thank you for participating!

Your task is to generate the funniest captions possible for a cartoon.

A caption involves language that enhances the humor of the cartoon and typically ranges from a few words to a few sentences long.

[Task Introduction]

[First, you will / First, you will / You will] be shown a cartoon and will be asked to generate as many humorous captions as you can.

[Then, you will be asked to select your best captions. Specifically, you will be asked to review the captions that you have generated and submit them for evaluation by an independent panel. You can submit as many captions as you would like with an opportunity to earn a bonus payment. /

Then, you will be asked to help us select the best captions. Specifically, you will be asked to review the captions that another participant in this study has generated and submit them for evaluation by an independent panel. You can submit as many captions as you would like with an opportunity to earn a bonus payment. /

You will submit these captions for evaluation by an independent panel. You can submit as many captions as you would like with an opportunity to earn a bonus payment.]

[Bonus Determination Procedure]

Bonus payments will be calculated as follows:

The captions you submit will be judged by an independent panel of Prolific participants on a 1-to-5 scale (1 = "Not at all humorous" to 5 = "Extremely humorous").

For every caption that you submit which receives an average rating of 4 or higher, you will earn a bonus payment of $0.25. However, for every caption that you submit which receives an

average rating below 4, you will lose $0.25 in bonus payment. Once all of the captions you submit have been rated, your total bonus payment will be calculated. Note: Any negative totals will result in a $0 bonus.

Only the captions that you submit are eligible to contribute to your bonus payment. Accordingly, whereas you can submit as many [of your captions / of the other participant's captions / captions] as you would like, you should only submit the captions which you think are likely to be rated favorably by the panel.

[The Quiz About the Task and Incentives]

- In this task, is there an opportunity to earn a bonus payment? (Yes / No)
- Whose captions will you be submitting to receive a bonus payment? (My captions / Someone else's captions)
- The captions you submit will be rated by a separate and independent panel of Prolific participants. (True / False)
- Only the captions you submit are eligible for a bonus payment. (True / False)
- If you submit captions that are rated less favorably by the independent panel, your bonus payment will be decreased. (True / False)

[Main Task, Stage 1: Idea Generation]

[Although the cartoon image is redacted due to copyright issues, it can be retrieved at the following URL:

https://web.archive.org/web/20200815180009/https://hbr.org/2013/11/strategic-humor-cartoons-from-the-december-2013-issue

The image depicts two businesspeople in a small boat, floating in water among drifting briefcases. It was used for a cartoon caption contest for Harvard Business Review (deadline: December 14, 2014).]

[Generate as many captions for this cartoon as you can.

At this stage don't worry about how funny each caption is—just try to generate as many captions as possible. /

Generate as many captions for this cartoon as you can.

At this stage don't worry about how funny each caption is—just try to generate as many captions as possible. /

Generate captions for this cartoon.

Recall that for every caption that you submit which receives an average rating of 4 or higher, you will earn a bonus payment of $0.25. However, for every caption that you submit which receives an average rating below 4, you will lose $0.25 in bonus payment.

Take as much time as you would like. Once you have entered all of the captions (and only the captions) you would like to submit for evaluation, please click the arrow button at the bottom of the page.]

Caption 1:

Caption 2:

Caption 3:

Caption 4:

Caption 5:

Caption 6:

Caption 7:

Caption 8:

Caption 9:

Caption 10:

[Main Task, Stage 2: Idea Selection, Only in the Self and Other Conditions]

Now, we'd like you to review the captions that [you / another participant] generated.

Please read each of the captions below and select the caption(s) you wish to submit for evaluation by the independent panel of Prolific participants.

You can also unselect any caption(s) by clicking on them again.

Recall that for every caption that you submit which receives an average rating of 4 or higher, you will earn a bonus payment of $0.25. However, for every caption that you submit which receives an average rating below 4, you will lose $0.25 in bonus payment.

Take as much time as you would like. Once you have selected all of the captions (and only the captions) you would like to submit for evaluation, please click the arrow button at the bottom of the page.

[Below, the captions appeared in buttons that participants could click (highlight) or unclick (unhighlight). All the captions initially generated (and only these captions) were shown as captions that could be selected and submitted. For example, if a participant in the Self condition generated three captions, all three captions were presented as three buttons, with Captions 4-10 not appearing as buttons. Figure S1. illustrates the analogous page for Study 1.]

[The cartoon image]

Caption 1: [the first caption that the participant (Self) or another participant (Other) generated]
Caption 2: [the second caption that the participant (Self) or another participant (Other) generated]
Caption 3: [the third caption that the participant (Self) or another participant (Other) generated]
Caption 4: [the fourth caption that the participant (Self) or another participant (Other) generated]
Caption 5: [the fifth caption that the participant (Self) or another participant (Other) generated]
Caption 6: [the sixth caption that the participant (Self) or another participant (Other) generated]
Caption 7: [the seventh caption that the participant (Self) or another participant (Other) generated]
Caption 8: [the eighth caption that the participant (Self) or another participant (Other) generated]
Caption 9: [the ninth caption that the participant (Self) or another participant (Other) generated]
Caption 10: [the tenth caption that the participant (Self) or another participant (Other) generated]

[Exploratory Items]

You submitted a total of [the number of the given participant's submitted captions] captions.

How many of the [the number of the given participant's submitted captions] captions do you think will receive an average rating of 4 or higher?

______ caption(s)

You were asked to select among [your / another participant's] captions. How large a bonus would you have earned if you were instead asked to select among [another participant's / your] captions?

(Much smaller bonus / Somewhat smaller bonus / Slightly smaller bonus / Same-size bonus / Slightly larger bonus / Somewhat larger bonus / Much larger bonus)

[Demographic Information]

- How old are you? ____
- What is your gender? (Male / Female / Other: ______)
- What is your race or ethnicity? (White / Hispanic, Latino, or Spanish / Black or African American / Asian / American Indian or Alaska Native / Middle Eastern or North African / Native Hawaiian or Other Pacific Islander / Some other race or ethnicity)
- What is the highest level of education you have completed? (Did not complete high school / High school graduate / Some college, no degree / Associate's degree / Bachelor's degree / Mater's, Professional, Doctorate degree)
- Approximately, how much is your annual household income? ($0 - $10,000 / $10,001 - $20,000 / … / $140,001 or Above)

**Section 5.** Study 4 Design and Materials

Study Design: 2 (Output Source: Self vs. Other) × 2 (Exemplars: Present vs. Absent)

[Below, the alternative wordings separated by "/" between square brackets show the wordings used for each of the two Output Source conditions: (1) Self and (2) Other conditions (respectively).]

[Study Introduction]

Thank you for participating!

In this study, your task is to generate the funniest captions possible for a cartoon.

A caption involves language that enhances the humor of the cartoon and typically ranges from a few words to a few sentences long.

[Task Introduction]

First, you will be shown a cartoon and will be asked to generate as many humorous captions as you can (up to 10).

Then, you will be asked to [select your best captions / help us select the best captions]. Specifically, you will be asked to review the captions that [you have / another participant in this study has] generated and submit them for evaluation by an independent panel. You can submit as many captions as you would like with an opportunity to earn a bonus payment.

[Bonus Determination Procedure]

Bonus payments will be calculated as follows:

The captions you submit will be judged by an independent panel of Prolific participants on the following 5-point scale:

1 = Not at all humorous
2 = Slightly humorous
3 = Somewhat humorous
4 = Very humorous
5 = Extremely humorous

For every caption that you submit which receives an average rating of 3 or higher, you will earn a bonus payment of $0.25. However, for every caption that you submit which receives an average rating below 3, you will lose $0.25 in bonus payment. Once all of the captions you submit have been rated, your total bonus payment will be calculated. Note: Any negative totals will result in a $0 bonus.

Only the captions that you submit are eligible to contribute to your bonus payment. Accordingly, whereas you can submit as many of [your / the other participant's] captions as you would like, you should only submit the captions which you think are likely to be rated favorably by the panel.

[The Quiz About the Task and Incentives]

- In this task, is there an opportunity to earn a bonus payment? (Yes / No)
- Whose captions will you be submitting to receive a bonus payment? (My captions / Someone else's captions)
- The captions you submit will be rated by a separate and independent panel of Prolific participants. (True / False)
- Only the captions you submit are eligible for a bonus payment. (True / False)
- If you submit captions that are rated less favorably by the independent panel, your bonus payment will be decreased. (True / False)

[Main Task, Stage 1: Idea Generation]

[Although the cartoon image is redacted due to copyright issues, it can be retrieved at the following URL:

https://web.archive.org/web/20200815180009/https://hbr.org/2013/11/strategic-humor-cartoons-from-the-december-2013-issue

The image depicts two businesspeople in a small boat, floating in water among drifting briefcases. It was used for a cartoon caption contest for Harvard Business Review (deadline: December 14, 2014).]

Generate as many captions for this cartoon as you can.

At this stage don't worry about how funny each caption is—just try to generate as many captions as possible (up to 10).

Caption 1:

Caption 2:

Caption 3:

Caption 4:

Caption 5:

Caption 6:

Caption 7:

Caption 8:

Caption 9:

Caption 10:

[Present Exemplars, Only in the Two Exemplars Present Conditions]

[In this section, three captions were randomly selected from a set of 49 captions with the mean humor rating of 3 or higher in a previous study. Each of these three captions was presented along with the cartoon. See Figure S2 for examples.]

Now, your task is to select captions.

To help you with selecting the best captions, we would like to show you some captions generated by other participants in a previous study.

All of the captions shown below received an average humor rating of 3 or higher.

[Three exemplar captions were presented sequentially on the same page; see Figure S2 for examples of these exemplar captions.]

[Main Task, Stage 2: Idea Selection]

Now, we'd like you to review the captions that [you / another participant] generated.

Please review each of the captions below and select the caption(s) you wish to submit for evaluation by the independent panel of Prolific participants.

You can also unselect any caption(s) by clicking on them again.

Recall that for every caption that you submit which receives an average rating of 3 or higher, you will earn a bonus payment of $0.25. However, for every caption that you submit which receives an average rating below 3, you will lose $0.25 in bonus payment.

Take as much time as you would like. Once you have selected all of the captions (and only the captions) you would like to submit for evaluation, please click the arrow button at the bottom of the page.

[Below, the captions appeared in buttons that participants could click (highlight) or unclick (unhighlight). All the captions initially generated (and only these captions) were shown as captions that could be selected and submitted. For example, if a participant in the Self condition generated three captions, all three captions were presented as three buttons, with Captions 4-10 not appearing as buttons. Figure S1. illustrates the analogous page for Study 1.]

[The cartoon image]

Caption 1: [the first caption that the participant (Self) or another participant (Other) generated]
Caption 2: [the second caption that the participant (Self) or another participant (Other) generated]
Caption 3: [the third caption that the participant (Self) or another participant (Other) generated]
Caption 4: [the fourth caption that the participant (Self) or another participant (Other) generated]

Caption 5: [the fifth caption that the participant (Self) or another participant (Other) generated]
Caption 6: [the sixth caption that the participant (Self) or another participant (Other) generated]
Caption 7: [the seventh caption that the participant (Self) or another participant (Other) generated]
Caption 8: [the eighth caption that the participant (Self) or another participant (Other) generated]
Caption 9: [the ninth caption that the participant (Self) or another participant (Other) generated]
Caption 10: [the tenth caption that the participant (Self) or another participant (Other) generated]

[Exploratory Items]

You submitted a total of [the number of the given participant's submitted captions] captions.

How many of the [the number of the given participant's submitted captions] captions do you think will receive an average rating of 3 or higher?

______ caption(s)

You were asked to select among [your / another participant's] captions. How large a bonus would you have earned if you were instead asked to select among [another participant's / your] captions?

(Much smaller bonus / Somewhat smaller bonus / Slightly smaller bonus / Same-size bonus / Slightly larger bonus / Somewhat larger bonus / Much larger bonus)

[Demographic Information]

- How old are you? ____
- What is your gender? (Male / Female / Other: ______)
- What is your race or ethnicity? (White / Hispanic, Latino, or Spanish / Black or African American / Asian / American Indian or Alaska Native / Middle Eastern or North African / Native Hawaiian or Other Pacific Islander / Some other race or ethnicity)
- What is the highest level of education you have completed? (Did not complete high school / High school graduate / Some college, no degree / Associate's degree / Bachelor's degree / Mater's, Professional, Doctorate degree)
- Approximately, how much is your annual household income? ($0 - $10,000 / $10,001 - $20,000 / … / $140,001 or Above)

**Figure S2.** Examples of the Exemplar Captions in Study 4

[The cartoon image]

We should have stuck to Zoom meetings.

[The cartoon image]

I guess you could say our assets are liquid now.

**Section 6.** Study 5 Design and Materials

Study Design: 2 (Output Source: Self vs. Other) × 2 (Intervention: Present vs. Absent)

[Below, the alternative wordings separated by "/" between square brackets show the wordings used for each of the two Output Source conditions: (1) Self and (2) Other conditions (respectively).]

[Study Introduction]

Thank you for participating!

In this study, your task is to generate the funniest captions possible for a cartoon.

A caption involves language that enhances the humor of the cartoon and typically ranges from a few words to a few sentences long.

[Task Introduction]

First, you will be shown a cartoon and will be asked to generate as many humorous captions as you can (up to 10).

Then, you will be asked to [select your best captions / help us select the best captions]. Specifically, you will be asked to review the captions that [you have / another participant in this study has] generated and submit them for evaluation by an independent panel. You can submit as many captions as you would like with an opportunity to earn a bonus payment.

[Bonus Determination Procedure]

Bonus payments will be calculated as follows:

The captions you submit will be judged by an independent panel of Prolific participants on the following 5-point scale:

1 = Not at all humorous
2 = Slightly humorous
3 = Somewhat humorous
4 = Very humorous
5 = Extremely humorous

For every caption that you submit which receives an average rating of 3 or higher, you will earn a bonus payment of $0.25. However, for every caption that you submit which receives an average rating below 3, you will lose $0.25 in bonus payment. Once all of the captions you submit have been rated, your total bonus payment will be calculated. Note: Any negative totals will result in a $0 bonus.

Only the captions that you submit are eligible to contribute to your bonus payment. Accordingly, whereas you can submit as many of [your / the other participant's] captions as you would like, you should only submit the captions which you think are likely to be rated favorably by the panel.

[The Quiz About the Task and Incentives]

- In this task, is there an opportunity to earn a bonus payment? (Yes / No)
- Whose captions will you be submitting to receive a bonus payment? (My captions / Someone else's captions)
- The captions you submit will be rated by a separate and independent panel of Prolific participants. (True / False)
- Only the captions you submit are eligible for a bonus payment. (True / False)
- If you submit captions that are rated less favorably by the independent panel, your bonus payment will be decreased. (True / False)

[Main Task, Stage 1: Idea Generation]

[Although the cartoon image is redacted due to copyright issues, it can be retrieved at the following URL:

https://web.archive.org/web/20200815180009/https://hbr.org/2013/11/strategic-humor-cartoons-from-the-december-2013-issue

The image depicts two businesspeople in a small boat, floating in water among drifting briefcases. It was used for a cartoon caption contest for Harvard Business Review (deadline: December 14, 2014).]

Generate as many captions for this cartoon as you can.

At this stage don't worry about how funny each caption is—just try to generate as many captions as possible (up to 10).

Caption 1:

Caption 2:

Caption 3:

Caption 4:

Caption 5:

Caption 6:

Caption 7:

Caption 8:

Caption 9:

Caption 10:

[Tell Participants About CEE, Only in the Two Intervention Present Conditions]

Now, your task is to select captions:

<< A note about selecting captions >>

Previous research has found that people tend to be overly optimistic when evaluating the quality of their own ideas.

In some cases, it can be beneficial to hand off your ideas to someone else in order to select the very best ones.

In your opinion, what is this research saying?

- People are not critical enough when evaluating their own ideas.
- People are too critical when evaluating their own ideas.
- I don't know.
  [If participants did not give the incorrect answer ("People are not critical enough…"), they could not advance to the next page and had to keep trying until they give the correct answer.]

[Main Task, Stage 2: Idea Selection]

Now, we'd like you to review the captions that [you / another participant] generated.

Please read each of the captions below and select the caption(s) you wish to submit for evaluation by the independent panel of Prolific participants.

You can also unselect any caption(s) by clicking on them again.

Recall that for every caption that you submit which receives an average rating of 3 or higher, you will earn a bonus payment of $0.25. However, for every caption that you submit which receives an average rating below 3, you will lose $0.25 in bonus payment.

Take as much time as you would like. Once you have selected all of the captions (and only the captions) you would like to submit for evaluation, please click the arrow button at the bottom of the page.

[Below, the captions appeared in buttons that participants could click (highlight) or unclick (unhighlight). All the captions initially generated (and only these captions) were shown as captions that could be selected and submitted. For example, if a participant in the Self condition generated three captions, all three captions were presented as three buttons, with Captions 4-10 not appearing as buttons. Figure S1. illustrates the analogous page for Study 1.]

[The cartoon image]

Caption 1: [the first caption that the participant (Self) or another participant (Other) generated]
Caption 2: [the second caption that the participant (Self) or another participant (Other) generated]
Caption 3: [the third caption that the participant (Self) or another participant (Other) generated]
Caption 4: [the fourth caption that the participant (Self) or another participant (Other) generated]
Caption 5: [the fifth caption that the participant (Self) or another participant (Other) generated]
Caption 6: [the sixth caption that the participant (Self) or another participant (Other) generated]
Caption 7: [the seventh caption that the participant (Self) or another participant (Other) generated]
Caption 8: [the eighth caption that the participant (Self) or another participant (Other) generated]
Caption 9: [the ninth caption that the participant (Self) or another participant (Other) generated]
Caption 10: [the tenth caption that the participant (Self) or another participant (Other) generated]

[Exploratory Items]

You submitted a total of [the number of the given participant's submitted captions] captions.

How many of the [the number of the given participant's submitted captions] captions do you think will receive an average rating of 3 or higher?

______ caption(s)

You were asked to select among [captions you generated / another participant's captions]. How large a bonus would you have earned if you were instead asked to select among [captions another participant generated / your captions]?

(Much smaller bonus / Somewhat smaller bonus / Slightly smaller bonus / Same-size bonus / Slightly larger bonus / Somewhat larger bonus / Much larger bonus)

[Demographic Information]

- How old are you? ____
- What is your gender? (Male / Female / Other: ______)
- What is your race or ethnicity? (White / Hispanic, Latino, or Spanish / Black or African American / Asian / American Indian or Alaska Native / Middle Eastern or North African / Native Hawaiian or Other Pacific Islander / Some other race or ethnicity)
- What is the highest level of education you have completed? (Did not complete high school / High school graduate / Some college, no degree / Associate's degree / Bachelor's degree / Mater's, Professional, Doctorate degree)
- Approximately, how much is your annual household income? ($0 - $10,000 / $10,001 - $20,000 / … / $140,001 or Above)

**Section 7.** Study 6 Design and Materials

Study Design: One Cell (Self Only)

[Study Introduction]

Thank you for participating!

Your task is to select funny captions for a cartoon.

A caption involves language that enhances the humor of the cartoon and typically ranges from a few words to a few sentences long.

[Task Introduction]

Previously, participants in our study generated humorous captions for the cartoon below.

[Although the cartoon image is redacted due to copyright issues, it can be retrieved at the following URL:

https://web.archive.org/web/20200815180009/https://hbr.org/2013/11/strategic-humor-cartoons-from-the-december-2013-issue

The image depicts two businesspeople in a small boat, floating in water among drifting briefcases. It was used for a cartoon caption contest for Harvard Business Review (deadline: December 14, 2014).]

You will be asked to help us select the best captions. Specifically, you will be asked to review the captions and submit them for evaluation by an independent panel. You can submit as many captions as you would like with an opportunity to earn a bonus payment.

[Bonus Determination Procedure]

Bonus payments will be calculated as follows:

The captions you submit will be judged by an independent panel of Prolific participants on the following 5-point scale:

1 = Not at all humorous
2 = Slightly humorous
3 = Somewhat humorous
4 = Very humorous
5 = Extremely humorous

For every caption that you submit which receives an average rating of 3 or higher, you will earn a bonus payment of $0.25. However, for every caption that you submit which receives an average rating below 3, you will lose $0.25 in bonus payment. Once all of the captions you submit have been rated, your total bonus payment will be calculated. Note: Any negative totals will result in a $0 bonus.

Only the captions that you submit are eligible to contribute to your bonus payment. Accordingly, whereas you can submit as many captions as you would like, you should only submit the captions which you think are likely to be rated favorably by the panel.

[The Quiz About the Task and Incentives]

- In this task, is there an opportunity to earn a bonus payment? (Yes / No)
- The captions you submit will be rated by a separate and independent panel of Prolific participants. (True / False)
- Only the captions you submit are eligible for a bonus payment. (True / False)
- If you submit captions that are rated less favorably by the independent panel, your bonus payment will be decreased. (True / False)

[Main Task, Idea Selection]

Please review each of the captions below and select the caption(s) you wish to submit for evaluation by the independent panel of Prolific participants. Since you were a participant in this study previously, it is possible that one or more of the captions that you will review were generated by you.

You can also unselect any caption(s) by clicking on them again.

Recall that for every caption that you submit which receives an average rating of 3 or higher, you will earn a bonus payment of $0.25. However, for every caption that you submit which receives an average rating below 3, you will lose $0.25 in bonus payment.

Take as much time as you would like. Once you have selected all of the captions (and only the captions) you would like to submit for evaluation, please click the arrow button at the bottom of the page.

[Below, the captions appeared in buttons that participants could click (highlight) or unclick (unhighlight). All the captions that participants had generated while participating in their previous study at t1 (i.e., six or more months prior and in the caption generation stage of the study), and only these captions, were shown as captions that could be selected and submitted. For example, if a participant generated three captions in the caption generation stage of their earlier study, all three captions were presented as three buttons, while Captions 4-10 would not appear as buttons. Figure S1. illustrates the analogous page for Study 1.]

[The cartoon image]

Caption 1: [the first caption that the participant (Self) or another participant (Other) generated]
Caption 2: [the second caption that the participant (Self) or another participant (Other) generated]
Caption 3: [the third caption that the participant (Self) or another participant (Other) generated]
Caption 4: [the fourth caption that the participant (Self) or another participant (Other) generated]
Caption 5: [the fifth caption that the participant (Self) or another participant (Other) generated]
Caption 6: [the sixth caption that the participant (Self) or another participant (Other) generated]
Caption 7: [the seventh caption that the participant (Self) or another participant (Other) generated]
Caption 8: [the eighth caption that the participant (Self) or another participant (Other) generated]
Caption 9: [the ninth caption that the participant (Self) or another participant (Other) generated]
Caption 10: [the tenth caption that the participant (Self) or another participant (Other) generated]

[Exploratory Items]

You submitted a total of [the number of the given participant's submitted captions] captions.

How many of the [the number of the given participant's submitted captions] captions do you think will receive an average rating of 3 or higher?

______ caption(s)

[Demographic Information]

- How old are you? ____
- What is your gender? (Male / Female / Other: ______)
- What is your race or ethnicity? (White / Hispanic, Latino, or Spanish / Black or African American / Asian / American Indian or Alaska Native / Middle Eastern or North African / Native Hawaiian or Other Pacific Islander / Some other race or ethnicity)
- What is the highest level of education you have completed? (Did not complete high school / High school graduate / Some college, no degree / Associate's degree / Bachelor's degree / Mater's, Professional, Doctorate degree)
- Approximately, how much is your annual household income? ($0 - $10,000 / $10,001 - $20,000 / … / $140,001 or Above)